\documentstyle[12pt,aasms4,epsf]{article}
\def\lsim{~\rlap{$<$}{\lower 1.0ex\hbox{$\sim$}}}
\def\gsim{~\rlap{$>$}{\lower 1.0ex\hbox{$\sim$}}}
\def\void#1{{}}

\newcommand{\h}{~h$^{-1}$ Mpc}
\newcommand{\etal}{{et al.~}}
\newcommand\kms{km~s$^{-1}$}
\newcommand\MB{$M_{B}$~}

\newcommand\eg{{e.g.\/}\rm,\ }
\input{epsf}
  \righthead{Properties of Very Luminous Galaxies} 
   \slugcomment{The Astronomical Journal, in press} 

\begin{document}

\title{Properties of Very Luminous Galaxies}
\author{A. Cappi}
\affil{Osservatorio Astronomico di Bologna, via Zamboni 33, I-40126,
   Bologna, Italy; cappi@astbo3.bo.astro.it}
\author{L.N. da Costa\altaffilmark{1}}
 \affil{European Southern Observatory, Karl-Schwarzschild-Str.2,
   D-85748 Garching bei M\"unchen, Germany; ldacosta@eso.org}  
\author{C.Benoist}
\affil{European Southern Observatory, Karl-Schwarzschild-Str.2,
   D-85748 Garching bei M\"unchen, Germany; cbenoist@eso.org}
\author{S. Maurogordato}
 \affil{CNRS; Observatoire de Nice, B4229, Le Mont--Gros, 06304 
Nice Cedex 4, France; maurogor@faure.obs-nice.fr}
\author{P. S. Pellegrini}
 \affil{Departamento de Astronomia CNPq/Observat\'orio Nacional, 
rua General Jos\'{e} Cristino 77, Rio de Janeiro, R.J. 20921 Brazil;
 pssp@dans.on.br}

\altaffiltext{1}{Departamento de Astronomia CNPq/Observat\'orio Nacional, 
rua General Jos\'{e} Cristino 77, Rio de Janeiro, R.J. 20921 Brazil}

\begin{abstract}
Recent analysis of the SSRS2 data based on cell--counts and two--point
correlation function has shown that very luminous galaxies are much
more strongly clustered than fainter galaxies. In fact, the amplitude
of the correlation function of very luminous galaxies ($L > L^*$)
asymptotically approaches that of $R \ge 0$ clusters. In this paper we
investigate the properties of the most luminous galaxies,
with blue absolute magnitude $M_B \le -21$. We find that: 1)
the population mix is comparable to that in other ranges of absolute
magnitudes; 2) only a small fraction are located in bona fide clusters; 
3) the bright galaxy--cluster cross--correlation
function is significantly higher on large scales than that measured
for fainter galaxies; 4) the correlation length of galaxies
brighter than \MB $ \sim -20.0$, expressed as a function of the mean
interparticle distance, appears to follow the universal dimensionless
correlation function found for clusters and radio galaxies; 5) a large
fraction of the bright galaxies are in interacting pairs, others show 
evidence for tidal distortions, while some appear to be surrounded by faint
satellite galaxies. 
We conclude that very luminous optical galaxies
differ from the normal population of galaxies both in the
clustering and other respects. We speculate
that this population is highly biased tracers of mass, being associated to 
dark halos with masses more comparable to clusters than typical loose
groups.

\end{abstract}

\keywords{
cosmology: observations --- dark matter --- 
galaxies:formation --- galaxies:fundamental parameters ---
galaxies: halos --- large-scale structure of universe
}


\section{Introduction}

Recently, the dependence of the clustering properties of galaxies on
luminosity was investigated by Benoist \etal (1996) and Willmer, da
Costa \& Pellegrini (1998) using volume-limited subsamples drawn from
the SSRS2 (da Costa \etal 1994, 1998).  Analysis based on the
two-point correlation function, in redshift and real space, and
counts-in-cells showed that luminous galaxies ($L > L^*$) are more
clustered than sub--L$^*$ galaxies. The effect is particularly strong
for the very luminous galaxies with \MB $\leq -21$ (hereafter VLGs),
for which the correlation length approaches that of $R \ge 0$ galaxy clusters.

This result might simply imply that the bright galaxies are
preferentially located in clusters. This would be the case, for
instance, if these galaxies were primarily cD galaxies, as suggested
by Hamilton (1988) in his analysis of the CfA1 catalog. Another
possibility is that very bright galaxies are located in loose groups
which are expected, on theoretical grounds (e.g. Hamilton \& Gott
1988), to have clustering properties intermediate between galaxies and
clusters, although no strong observational evidence currently exists
in support to this conjecture. Previous attempts to measure the
correlation function of loose groups in the CfA and SSRS surveys with
a standard algorithm (Huchra \& Geller 1982) have led to the
conclusion that the amplitude is comparable to --or even lower than--
that of galaxies (Jing \& Zhang 1988, Maia \& da Costa 1990, Ramella
et al. 1990).  Progress has recently been made by Girardi et
al. (1998) analyzing a much larger group catalog derived from CfA2 and
SSRS2. These authors show compelling evidence
that groups are indeed more clustered than galaxies with a correlation
length of the order of 10~\h, yielding a relative bias of $\approx$
3. They also find evidence that the correlation amplitude increases
with luminosity (mass) of the group.  Therefore the clustering
properties of VLGs might also be explained if VLGs are in very rich
groups.

Alternatively, if VLGs are not preferentially in rich groups or poor
clusters, an intriguing possibility is that they are associated to
massive dark halos and that their large correlation length reflects
the relative bias of these halos to the underlying mass distribution.
Semi--analytical models of galaxy evolution such as those of
Kauffmann, White \& Guiderdoni (1993) are now able to make some
predictions about the luminosity function of galaxies in halos of
different mass, despite the uncertainties related to the
galaxy--dark halo connection. An apparent generic prediction is that
each halo would contain a dominant galaxy surrounded by fainter
galaxies (Kauffmann, Nusser \& Steinmetz 1997). If this is the case,
then it would imply that at least some cases of dark matter
concentrations can be detected and used to investigate the
halo--galaxy connection.

Here we investigate in more detail the characteristics of very luminous
galaxies (hereafter VLGs) and compare their clustering properties to
those of galaxy clusters. In section 2, we describe the properties of
this galaxy population from available data in the literature and
investigate the environment in which they reside. In section 3, we
examine their distribution relative to the large--scale structures and
their auto--correlation and VLG--cluster cross--correlation properties.
Our main conclusions are summarized in section 4.

\section{Characteristics of Very Luminous  Galaxies}

\subsection{VLG Sample}

The VLG sample was drawn from the complete magnitude--limited SSRS2
south catalog, which contains about 3400 galaxies with $m_B(0) \leq
15.5$ (da Costa et al. 1994, 1998), covering the region $ b \leq
-40^o$ and $ -40^o \leq \delta \leq -2.5^o$.  Absolute magnitudes were
estimated from luminosity distances, assuming $H_0 = 100$ and $q_0 =
0.5$. Velocities were corrected for the motion of Local Group and
apparent magnitudes corrected for K--dimming according to the
different morphological types. Our VLG sample consists of 113 galaxies
with $ M_B~\leq -21$ ($L > 4 L^*$). The redshift distribution for
these galaxies is shown in figure 1. We should point out that we
excluded from our sample a few cases of strongly interacting systems,
or of galaxies nearby bright stars, which had been preliminarily
selected in the catalog because their magnitude had been significantly
overestimated (a few VLGs we have included might still be
affected by this problem, e.g. VLG~108).  Together with the SSRS2
magnitude errors ($\sim 0.3$), there is some uncertainty in our VLG
catalog limited at $M \le -21$, but this has no impact in our
conclusions.

\subsection {Individual Properties}

In order to look for possible explanations for the strong clustering
exhibited by the VLGs (see Benoist \etal 1996 and discussion below) we
first examine the characteristics of the individual galaxies that
comprise this population. For this purpose, we have used the
morphological classification as available in the SSRS2 catalog (da Costa
et al. 1998) and have searched the NASA Extragalactic Database (NED) for
additional information on these galaxies.

Our first finding is that the morphological composition of our VLG
sample does not differ from
the whole SSRS2 sample. In the bright sample there are 39 (35\%)
early--type galaxies ($T < 0$) and 74 (65\%) late-types. These fractions
are consistent, for example, to the 35\% of early types found in the \MB
$ \leq -19.0$ volume--limited subsample, which does not show the
enhanced correlation exhibited by the VLGs. It is worth stressing here
that morphological classification of these bright galaxies --at least the
separation between early and late types-- is quite reliable (da Costa
et al. 1998). 

From the NED we also find that only four galaxies are cDs,
which, as mentioned, would be the most natural explanation for the
observed strong clustering, as in that case one would be measuring
the amplitude of the cluster--cluster correlation.

In order to illustrate the properties of VLG galaxies, we summarize in
table 1 the information gathered from the NED for all the VLG galaxies.
The table includes: in column (1) the identification number;
in column (2) the catalog name; in column (3)
the ESO name (Lauberts 1980), whenever available; in columns (4) and (5)
the B1950 right ascension and declination; in column (6) the apparent
magnitude as listed in the SSRS2 catalog; in column (7) the B-R color
(Lauberts \& Valentjin 1989; de Vaucouleurs et al. 1991), 
whenever available; in column (8) the infrared luminosity at $60 \mu$m
in solar units (with $H_0 = 100$~km~s$^{-1}$~Mpc$^{-1}$); in column (9)
the morphological type either from the SSRS2 catalog or from the literature
if more detailed classification is available; in column (10) the
heliocentric radial velocity;  in column (11)  an indication of other
properties such as infrared emission (ir), AGN (an), radio source (rs).

We call attention that for the 13 VLG galaxies with radial velocities
less than 10,000 \kms, for which the information should be the most
complete, Table 1 shows that late--spirals is the dominant morphological
type, often with bars and rings. Moreover, 6 out of the 13 nearby
VLGs are also detected by {\em IRAS}, typically with $L_{IR} \sim
10^{10}$ in solar luminosity units.

For the population as a whole we find that a large number show
other interesting characteristics: 1) there are 30 peculiar galaxies
(Arp and Madore 1987); 2) 42 galaxies are also in the IRAS Point Source
Catalog and/or in the Faint Source Catalog (Moshir et al. 1990); 3) we
find an apparent overabundance of barred and ring galaxies. Out of 26
galaxies with detailed morphological information 13 contain bar/ring
structure, while 7 are intermediate cases; 4) 14 galaxies are
radio--sources; 5) only 5 galaxies are AGNs (3 Seyferts and 2 Liners;
Maia \etal 1996, NED), showing that the VLGs have luminosities that
exceed some nearby AGN--like galaxies.

In the VLG sample, we find that 54 are included in the ESO catalog
for which $B$ and $R$ magnitudes are available. In figure 2 we compare
the color distribution of this subsample with that of the $M \le -19$
sample, renormalized to the total number of VLGs. While VLGs have
a slightly larger spread in color, the 2 distributions are similar.

The above results indicate that the sample of SSRS2 galaxies drawn 
on the basis of their large blue luminosity is ``special'' 
relative to galaxies of lower luminosity both in 
terms of their internal characteristics as well as their clustering 
properties.

\subsection {Relation VLGs--Clusters}

As suggested earlier, an obvious explanation for the large clustering
strength of bright galaxies would be that we are simply picking up
cluster members, thus measuring the amplitude of the cluster--cluster
correlation function (\eg Hamilton 1988). Although, as previously
mentioned, the VLGs are not cDs, they could still reside preferentially
in rich clusters. Even though the volume covered by the SSRS2 does not
include very rich clusters, we have examined the ACO catalog
(Abell, Corwin \& Olowin 1989), which can
be considered complete out to a distance of 200 \h, searching for VLGs
with a projected separation less than 1.5\h~ from cataloged $R \ge 0$
cluster centers and a radial velocity difference less than 1200 \kms~
with respect to the cluster mean velocity. We find only 12 galaxies
which can be considered as candidate cluster members. We have also
extended our search to the poor clusters in the volume, using the ACO
supplementary list and the Edinburgh--Durham cluster catalog (Nichol et
al. 1992), with a measured redshift. In this search we include  clusters
beyond the redshift limit of the bright galaxy sample. Using the same
criteria as that used for the rich clusters we find 7 additional
galaxies which could be associated with known poor clusters.  Therefore,
at best only 19 VLGs may reside in previously known rich and poor
clusters, out of which 16 are early--type galaxies. This result
confirms that very luminous galaxies are not preferentially in 
$R \ge 0$ or rich clusters.

In figure \ref{fig:3} we show the spatial distribution of the VLGs and
the ACO clusters in the region, in the declination range $-40^o \le
\delta \le -2.5^o$.  In order to have a preliminary understanding of the
VLG location relative to the large--scale structures, we have resorted
to a simple percolation analysis. Adopting a percolation parameter $s =
0.5$, which corresponds to a search radius $r_s = 12.6$ \h~ and to a
space density enhancement of $1.9$, we have detected 7 structures with
more than 3 members, containing about 42\% of the bright galaxies. Among
them there are two main structures. One is at a mean redshift $z \sim
0.055$. It consists of 20 members, out of which only 5 are possible
members of rich clusters, all belonging to the
Pisces--Cetus region. The second association, with 9 members, is at a
mean redshift $z \sim 0.054$, and none of its galaxies are found in rich
clusters.  Comparing their coordinates with maps of the large--scale
structures, based on the ACO cluster distribution (Tully 1986, Tully
\etal 1992), both these associations correspond to dense regions of the
Pisces--Cetus Supercluster.

As a final note, we point out that eliminating the VLGs which may be in
clusters does not significantly change the correlation function of the
sample (see section 3).

\subsection {Environment of VLGs}

As we have seen, there is no compelling evidence that VLGs reside
preferentially in clusters of galaxies. However, this does not fully
answer the more general question about the type of environment in
which these galaxies reside. 
Unfortunately, this is not an easy task because the
majority of the VLGs are at large distances, where fainter neighbors
would not be included in the SSRS2 sample because of its relative bright
magnitude limit. On the other hand, the use of radial velocity databases
is dangerous because of their unknown incompleteness.  Therefore, we
can only attempt a preliminary investigation of the question.

First of all, we have used the catalog of groups recently identified
within the SSRS2 sample by Girardi et al. (1998). This catalog
contains groups with mean positions out to 12,000 \kms, but includes
members out to 15,000 \kms.  With this procedure we should in
principle be able to detect groups including nearby VLGs: at 12,000
\kms~ we can detect a group with a VLG and 2 companions brighter than
$\sim -19.9$ or even fainter at smaller distance (there are 27 VLGs
within $V \le 12,000$\kms, and 45 VLGs within $V \le 15,000$\kms).
We find 14 VLGs in 13 loose groups; of these, 3 are in known clusters
(and have already been included in the 19 mentioned above), 3 are in
known groups --2 in Hickson compact groups 12 and 91 (see Hickson
1982), one in an Arp--Madore group--, and 8 are in new groups,
hereafter referred as SSRS2 groups. The richness of the 8 loose
groups (i.e. those which are not associated with rich clusters or compact
groups) is quite low; 4 groups have 3 members, 2 groups have 5 members,
and 1 has 7 members, including 2 VLGs.

The identification of groups in the total VLG sample is more
difficult, and is necessarily incomplete.  Using the whole SSRS2
catalog, we have selected galaxies with a projected separation of 1.5
\h~ and within $1200$ km/s from a very luminous galaxy. This is not a
real group finding algorithm but it suffices for our purposes. In this
way we have found 22 galaxies in group candidates with at least 3
members (including the VLG); most of them are known systems we have
already taken into account: only 4 are new candidate systems, which
have not been identified in the catalogs of clusters or groups.

In addition, we have also searched VLG galaxies in the compact group
catalog by Barton et al. (1996). We find two systems which include
nearby VLGs: their no.81 (which corresponds to HCG91) and no.88, which
corresponds to the core of A4038. Both these cases are also in the
SSRS2 group catalog (Girardi et al. 1998).

However, as we have stated the number of systems we can identify is
biased by the relatively bright magnitude limit of the SSRS2, which
could lead us to miss fainter galaxies. Therefore, we have also
searched the literature (\eg NED) to identify known systems of
galaxies from binaries to groups. Of course, one should be cautious in
interpreting the results because of the incompleteness of these data
sets.  From this search we find for example other 2 VLGs in southern
compact groups from the automated catalog by Prandoni, Iovino \&
MacGillivray (1994), 13 pairs and 4 triplets (2 pairs belonging to
systems previously found).

Combining all the above results we may conclude that out of the 113
VLGs we have identified 12 in rich clusters, 7 in poor clusters, 24 in
generic groups with at least 3 members and 11 in interacting pairs or
possible binaries, for a total of 54 VLGs in some type of galaxy
association.  In table 2, we list all VLGs that have been identified
with some system ranging from binaries to rich cluster.  In column (1)
we give the identification number, in column (2) the catalog name, in
column (3) the system type (binary, triplet, group, poor cluster, rich
cluster), with an eventual note on the VLG (cD, Seyfert, interacting),
in column (4) the system name (ACO or EDCC name for clusters, Hickson
no. for compact groups, Arp--Madore, ESO or Vorontsov--Velyaminov
no. for the others, indicating simply SSRS2 for groups listed only in
the Girardi et al. (1998) catalog or for candidate groups found by
percolation in this work), in column (5) the velocity dispersion of
the system as measured from the SSRS2, in column (6) the number of
galaxies used to estimate the velocity dispersion. As apparent
from table 2, there is generally only one VLG per system. Two exceptions
are A4008, which contains 4 VLGs, and the SSRS2 loose group
which includes VLG~37 and VLG~38.

From the above identified systems, extended X--ray emission,
consistent with the VLG position, was detected by ROSAT for galaxies
in A4038 and S0141. The object VLG 86, which is a Seyfert 1, was
detected as a point source in X by ROSAT (B\"ohringer, private
communication). At the position of VLG 65, ROSAT detected also X--ray
extended emission typical of a group or cluster. This is an
interesting object, because even if it is relatively nearby ($z \sim
0.03$), it is not found in any cluster or group catalog. We identified
it in the SSRS2 through percolation as a group of 3 galaxies
(including the VLG) and in a note in the MCG (Vorontsov--Velyaminov \&
Arhipova 1968) referring to this galaxy we read: ``Here is the main
member of a cluster, located at its end.  The other members are of
type E and S0 and are much fainter, but visibly spiral and [16-17
mag].''

The remaining 59 VLGs have not been associated to any particular known
system. However, after eye--examination of these galaxies using the
Digitized Sky Survey (DSS), we estimate that only about 16 galaxies
might be isolated. Among the other galaxies, about 20 have at least a
companion, and show clear distorted morphologies and/or evidence of
interaction, 5 appear to be surrounded by faint satellites and 18
might be in groups (often with members much fainter than the VLG, as
in the case of VLG~65), although usually no redshift information is
available. This is illustrated in figure 4 for four such cases: VLG~24
is surely not a typical case, being an elliptical galaxy presumably at
the center of a group; VLG~41 is probably a member of a triplet/small
group of galaxies; VLG~44 is clearly interacting with a large,
irregular galaxy; VLG~77 is a Liner, probably interacting with a small
satellite.

Another evidence that VLGs are unlikely to be isolated systems, comes
from the examination of the small subsample of nearby VLGs ($V <
10,000$ \kms, a distance at which a typical $M^* \sim -19.5$
galaxy is brighter than m=15.5 and is therefore included in the
SSRS2). Images for these galaxies are shown in figure 5. Most of
them are not isolated. For example, VLG~20 does not belong to any
known cluster or group, but we have identified around it (through the
percolation technique previously described) other 7 fainter SSRS2
galaxies, giving a velocity dispersion of $\sim 900$ km/s (see table
2).  Close inspection of the images for the only three nearby VLGs,
VLG~27, VLG~29, and VLG~61, which, from our analysis, would be
classified as ``isolated", show instead that they have companions:
VLG~27, a peculiar barred late--type spiral, seems to be surrounded by
satellites very close it as well as brighter galaxies at larger
distances; VLG~29 (listed in the Arp--Madore catalog of Southern
Peculiar galaxies) has a disk galaxy as a probable companion; VLG~61
has many nearby satellites.

From the above evidence we can conclude that VLGs are generally
found in environments of high local galaxy density, but not in rich
clusters. Moreover, both eye-inspection of the images of nearby VLGs
and their surroundings (within 1 \h~ at the VLG redshift), and the
number of members in SSRS2 loose groups found to include a VLG, show
no evidence that VLGs are preferentially in rich groups.

\void{
This is an
important point, as the clustering properties we will examine in the
next section suggest that VLGs are indeed associated to massive dark
halos.}

\section{Large--scale Distribution}

\subsection {Correlation function}

The original reason for focusing our attention in the sample of VLGs
was the strong bias showed by this population relative to fainter
galaxies (Benoist \etal 1996).  This is illustrated in figure
\ref{fig:test} where we show the correlation function for galaxies
brighter than $ M = -21$ within 168 \h.  For comparison, we also show
the fit to the cluster correlation function (Cappi \& Maurogordato
1992) determined for $R \ge 0$ ACO clusters (Abell et al. 1989), $
\xi_{cc} = (s/19.5)^{-1.8}$.

We remind that the correlation function of clusters depends on
their richness (see Mann et al. 1993, Croft et al. 1997) For $R \ge 1$
clusters Peacock \& West (1992) find $21 \pm 1.3$ \h, but some authors
claim that the correlation amplitude of Abell clusters is amplified by
projection effects, and give an estimate of $\sim 14$ \h~ for APM
clusters (Dalton et al. 1994), which would then be comparable to the
the value found for VLGs.

From the figure we see that very bright galaxies have a correlation
length comparable to that measured for poor clusters. The correlation
length of \MB $\leq -21$ galaxies is $r_0 = 16 \pm 2$ \h, while the
zero--crossing of the correlation function is beyond $\sim 40$ \h.

In Benoist \etal (1996), it was possible to prove that the dependence of
the correlation amplitude on luminosity was real,  by considering
different luminosity-limited samples within the same volume.
Unfortunately, the same direct test is not possible for the $M \le -21$
sample, because of the relatively small number of galaxies. Therefore,
one might argue that the large correlation amplitude is due to sampling
effects, caused either by virialized systems in the volume considered or
by fluctuations of the mean background density, on scales comparable to
the sample size; in fact, examining figure \ref{fig:3}, where we show
the spatial distribution of the bright galaxies, in the declination
range $-40^o \le \delta \le -2.5^o$, at least two concentrations can be
seen at large distances, at $\alpha \sim 23^h 30^m$ and at $\alpha \sim
1^h$, corresponding to large--scale structures seen
in redshift space, as discussed in section 2.3.

There are, however, two lines of argument showing that our
results are not a consequence of sampling effects. The first, based
on circumstantial evidences, is that: 1) as we have seen the bright
galaxies seem to have peculiar characteristics; 2) their clustering
properties extend a trend already visible for fainter galaxies (see
figure 5 of Benoist et al. 1996); 3) similar results were found by
Hamilton (1988) in his analysis of the CfA1 catalog, i.e. probing a
different region of the sky, and by Park et al. (1994) from the
analysis of the CfA2 power spectrum. A more direct argument is based
on the following facts: 1) the correlation amplitude does not vary
significantly if one considers sub-samples defined by removing the
main concentrations; 2) by using a smaller volume; 3) or by splitting
the sample into two ranges of right ascension, probing different
structures.

Therefore, the large correlation amplitude of very bright galaxies is
probably a genuine property. We also point out that the amplitude of
the correlation is not affected by removing galaxies identified in the
previous section as possibly belonging to rich clusters.  It is also
important to emphasize that although a large fraction of the VLGs may
be in systems of galaxies, currently there is no evidence for the
amplitude of the SSRS2 group--group correlation function (Girardi et
al. 1998) to be as large as that obtained for the VLGs. On the
basis of the evidence presented in the previous section, VLGs are not
preferentially in clusters, and probably not in rich groups (this last
claim is obviously based on our analysis of nearby VLGs). We therefore
suggest that VLGs are a highly biased population, possibly associated
with dark halos with masses comparable to clusters.

\subsection{VLG--Cluster cross--correlation function}

An additional evidence in support of the above conjecture is the
VLG-cluster cross-correlation function shown in figure
\ref{fig:cross}a. Despite the small number of clusters (we have found
only 28 $R \ge 0$ ACO clusters within the volume considered), we find
$\xi_{gc} = \left( r / 16 \pm 2 \right) ^{-1.7 \pm 0.4}$. For
comparison we also show the galaxy--cluster cross-correlation
determined by Lilje \& Efstathiou (1988), $\xi_{gc} = \left( r / 8.8
\right) ^{-2.2}$, based on the Lick counts of galaxies.  While the
slopes differ at the $1 \sigma$ level, $r_0$ for bright galaxies is
significantly larger, by a factor of about 2, even though the
uncertainties are large (for example, Seldner \& Peebles 1977 find a
significantly larger amplitude for the galaxy--cluster
cross--correlation than Lilje \& Efstathiou 1988).

An estimate of the relative bias between the VLGs and clusters can be
computed from the ratio $\alpha = J_3 (gc) / J_3 (cc)$, where $J_3 (s) =
\int _0 ^s x^2 \xi(s) ds$, $J_3 (gc)$ is the integral of the
galaxy--cluster cross--correlation, and $J_3 (cc)$ is the integral of
the cluster correlation function (see Alimi \etal 1988). The dependence
of $\alpha$ on separation $s$ is shown in figure \ref{fig:cross}b from
which we determine that galaxies with \MB $ \leq -21$ are less clustered
by a factor of $b_{21}/b_{cc} \sim 0.8$ relative to clusters, which
appears to be constant in the range 5--50 \h. This high bias is
comparable to that measured for radio--galaxies (Peacock \& Nicholson
1991, Mo, Peacock \& Xia 1993, Loan, Wall \& Lahav 1997), and which has been
explained as being due to the fact that radio--galaxies reside
preferentially in clusters. However, as we have seen the VLGs are
neither radio-sources nor are located in rich clusters. Furthermore,
while radio--galaxies are representative of early--type galaxies, our
results show that late--type galaxies also contribute to the effect.

\subsection{Dimensionless Correlation Function}

In order to further compare the clustering properties of the bright
galaxies with those of clusters and other systems we use the scaling
relation proposed by Szalay \& Schramm (1985), who noted that the
cluster correlation amplitude increases with their mean spatial
separation.  As discussed by Bahcall \& West (1992) the ``universal
dimensionless correlation function"

\begin{equation}
A_i = r_0 ^\gamma \sim (\alpha d_i)^{1.8}
\label{eq:unicor}
\end{equation}

\noindent where $d_i = n^{-1/3}$ is the mean inter-particle distance
and $\alpha \sim 0.5$, seems to hold for a number of systems up to
superclusters, but not for normal galaxies. However, our results
suggest that in this respect $M \le -21$ galaxies should differ from
typical $M^*$ galaxies, as their mean density is much lower ($n \sim 6
\times 10^{-5}$~h$^3$~Mpc$^{-3}$ in the SSRS2), and they are more
strongly correlated.

In figure \ref{fig:universal} we show the correlation length $r_0$ for
different volume-limited subsamples of the SSRS2, as a function of their
mean inter-particle distance $d_i = n^{-1/3}$, where $n$ is the mean
density of the subsample. Filled hexagons refer to SSRS2 volume--limited
subsamples: each subsample includes galaxies in a one magnitude range,
from $-17 \le M_B \le -16$ (lowest point) to $-21 \le M_B \le -22$, with
a step of 0.5 magnitudes (for details, see Benoist \etal 1996).

Recalling that for the SSRS2, \MB$^* \sim -19.5$, it is clear from
figure \ref{fig:universal} that faint galaxies ($L < L^*$) show a
significant departure from the relation, as one could expect {\em a
priori} (see Peebles 1993), whereas brighter galaxies ($L > L^*$)
approach it asymptotically, which is a much less trivial
result. This is in fact the first time that it is possible to cover a
sufficiently large range of luminosities to allow a comparison of the
scaling relation of galaxies and clusters.  Galaxies brighter than \MB
$=-20$ approximately follow a scaling relation with $\alpha \sim 0.6$,
slightly higher but, given the uncertainties, consistent with the
value of $\alpha \sim 0.4$ determined by Bahcall \& West (1992) for a
variety of systems.
 
Szalay \& Schramm (1985; see also Luo \& Schramm 1992) argued that the
universal dimensionless correlation function followed by clusters could
be explained if structures formed by a truly scale--invariant process,
leading to a fractal--like distribution, while the observed deviation of
galaxies from this relation could be explained by non--linear
gravitational clustering at small scales. We take a different point of
view and argue that the observed behavior is a clear evidence for
biasing.  First, because the results of Benoist \etal (1996) show that
the correlation length does not depend on the depth of the sample as
expected for fractals (see also Cappi et al. 1998). 
Second, the clear trend of very bright galaxies
to lie close to the relation obeyed by clusters suggests a continuity of
the clustering properties of bright galaxies and clusters. This would be
a natural consequence of different biasing amplitudes for systems with
different mass scales (e.g. Peebles 1993).

Our results show that the universal dimensionless correlation function
can be explained as a consequence of the increase of the correlation
amplitude with the mass (and therefore rarity) of the system, as
expected if the distribution of luminous objects is biased with respect
to the underlying matter distribution (Kaiser 1984; Bardeen \etal 1986).
The $r_0 - d$ relation should thus be regarded as an empirical tool to
compare the biasing of different classes of objects (see for example the
more rigorous approach of Bernardeau \& Schaeffer 1992).

\section{Discussion and Conclusions}

From our analysis of the SSRS2, we have shown that very luminous
galaxies (\MB $ \leq -21$ or $L \gtrsim 4L^*$) are strongly correlated
($r_0 = 16 \pm 2$ \h) and the zero--crossing of the correlation
function is beyond 40\h.  Furthermore, we find that luminous galaxies
with $L > L^*$ appear to follow a universal dimensionless correlation
similar to that of galaxy clusters, which implies a common biasing
mechanism.

Independently of their clustering properties we have found that $L
\gtrsim 4L^*$ galaxies seem to form a special population. They are
observed in locally high galaxy density regions, but are not
predominantly in clusters nor apparently in rich groups.  Instead,
they tend to be in interacting pairs, some exhibit tidal distortions,
while others have faint satellite galaxies in their surroundings.
VLGs resemble another class of galaxies: the ultra--luminous infrared
galaxies, where the high luminosity appears to be triggered by
interaction with other galaxies (e.g. Melnick \& Mirabel 1990;
Clements et al. 1996; Duc et al. 1997).  Is there an evolutionary link
between these systems? Is it possible that a massive galaxy undergoes
a period of intense star formation, shrouded by dust, and becomes
finally visible as a VLG in the optical?  A simple test of this
hypothesis will be possible with a large and complete sample of
infrared ultra--luminous galaxies (which should have a higher
correlation amplitude than normal IRAS galaxies).

The fact that the brightest galaxies in the sample are not found
in clusters but have clustering properties similar to $R=0$ clusters
suggests that they may reside within massive dark halos. We cannot
completely exclude that VLGs are preferentially in rich groups, which
should be more strongly clustered than loose groups as a
whole, because most VLGs are at large distances, where fainter
members are not included in the SSRS2. However, as discussed above, at
the present time we find no evidence for that, as
most groups containing a nearby VLG are of low richness. While many
VLGs have companions, the total luminosity of these systems appear to
be far less than that typical of clusters. If this is the case, it
implies that there are large variations in the M/L ratio and/or in the
luminosity function of galaxies forming in dark halos. Further
observational work is clearly needed to fully characterize this galaxy
population. Clarifying the nature of these objects may contribute to
our understanding of the halo--galaxy connection and the mechanisms
responsible for galaxy evolution.

\acknowledgements 

This research has made use of the NASA / IPAC Extragalactic Database
(NED) which is operated by the Jet Propulsion Laboratory, California
Institute of Technology, under contract with the National Aeronautics
and Space Administration. The Digitized Sky Surveys were produced at
the Space Telescope Science Institute under U.S. Government grant NAG
W-2166. The images of these surveys are based on photographic data
obtained using the Oschin Schmidt Telescope on Palomar Mountain and
the UK Schmidt Telescope. The plates were processed into the present
compressed digital form with the permission of these institutions.
LNdC thanks his SSRS2 collaborators for allowing to use the data prior
to publication and the hospitality of the Institut d'Astrophysique and
the Observatoire de Paris--Meudon. AC thanks the hospitality of the
Observatoire de Nice. We wish also to thank H. B\"ohringer for
his list of X--ray detection in the fields of very luminous galaxies,
and the referee M. Ramella for his useful comments.

\vfill
\newpage

%
%
{\bf Table captions}

Table 1 --- List of the VLGs galaxies ($M_B \le -21$) in the SSRS2;
ir=detected by IRAS, rs=radiosource, an=AGN.

Table 2 --- List of VLGs in known systems of galaxies.

\vfill
\newpage 

{\bf Figure captions}

\figcaption[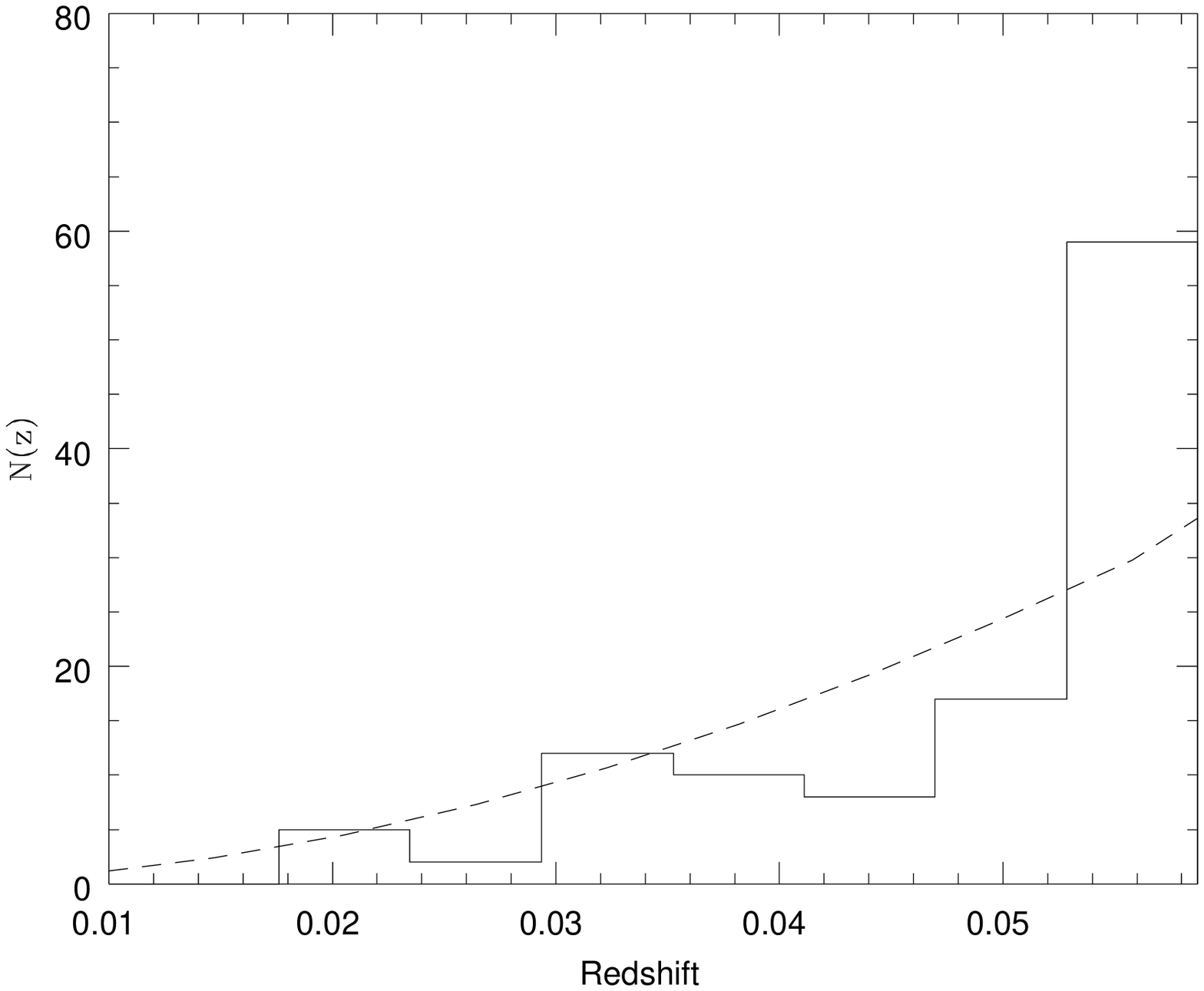] 
{Redshift distribution of VLG galaxies.}

\figcaption[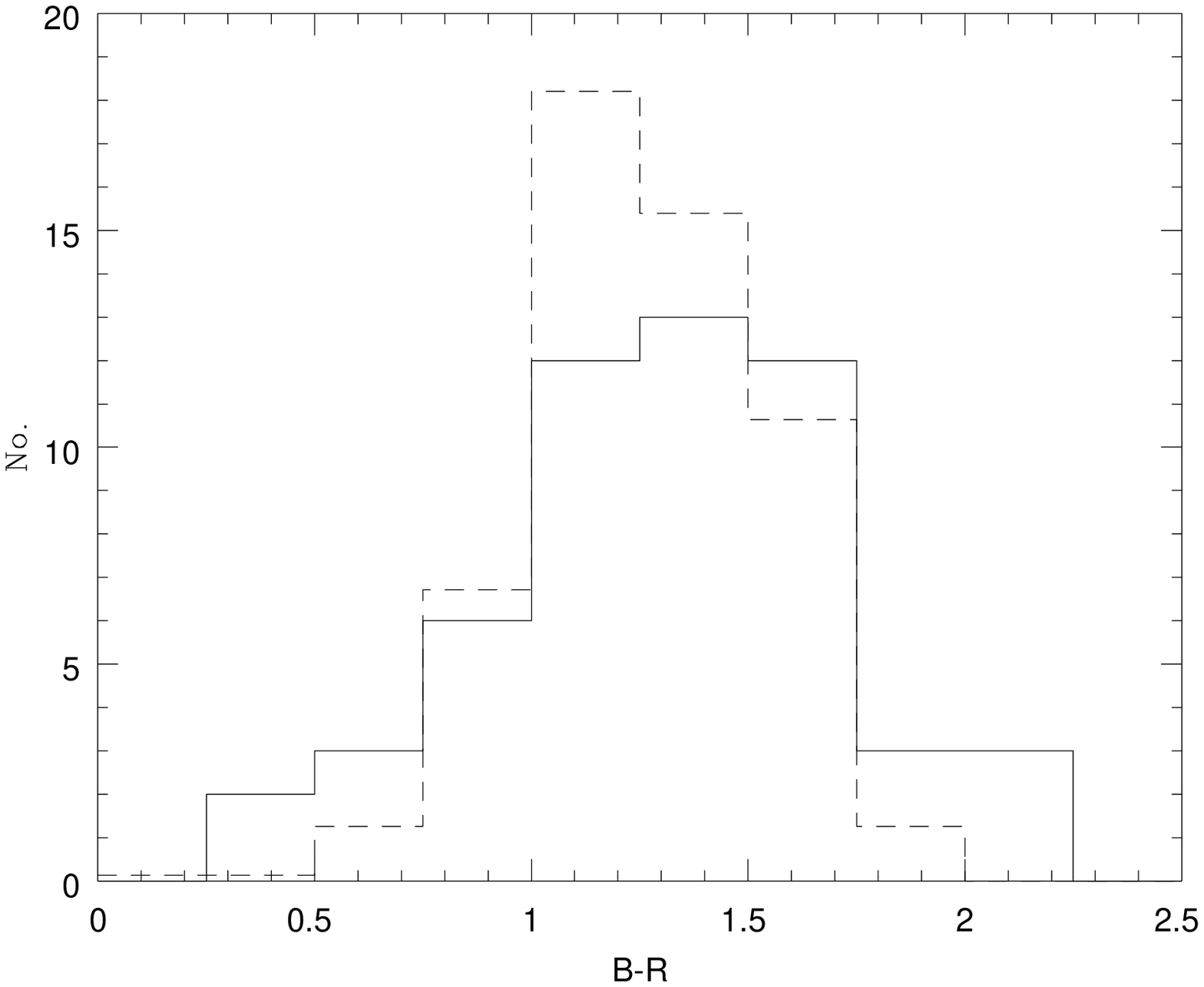]{
Histogram Comparison of the $B-R$ color distribution
of VLGs (solid line) and $M \le -19$ galaxies
(dashed line; renormalized to the total no. of VLGs).}

\figcaption[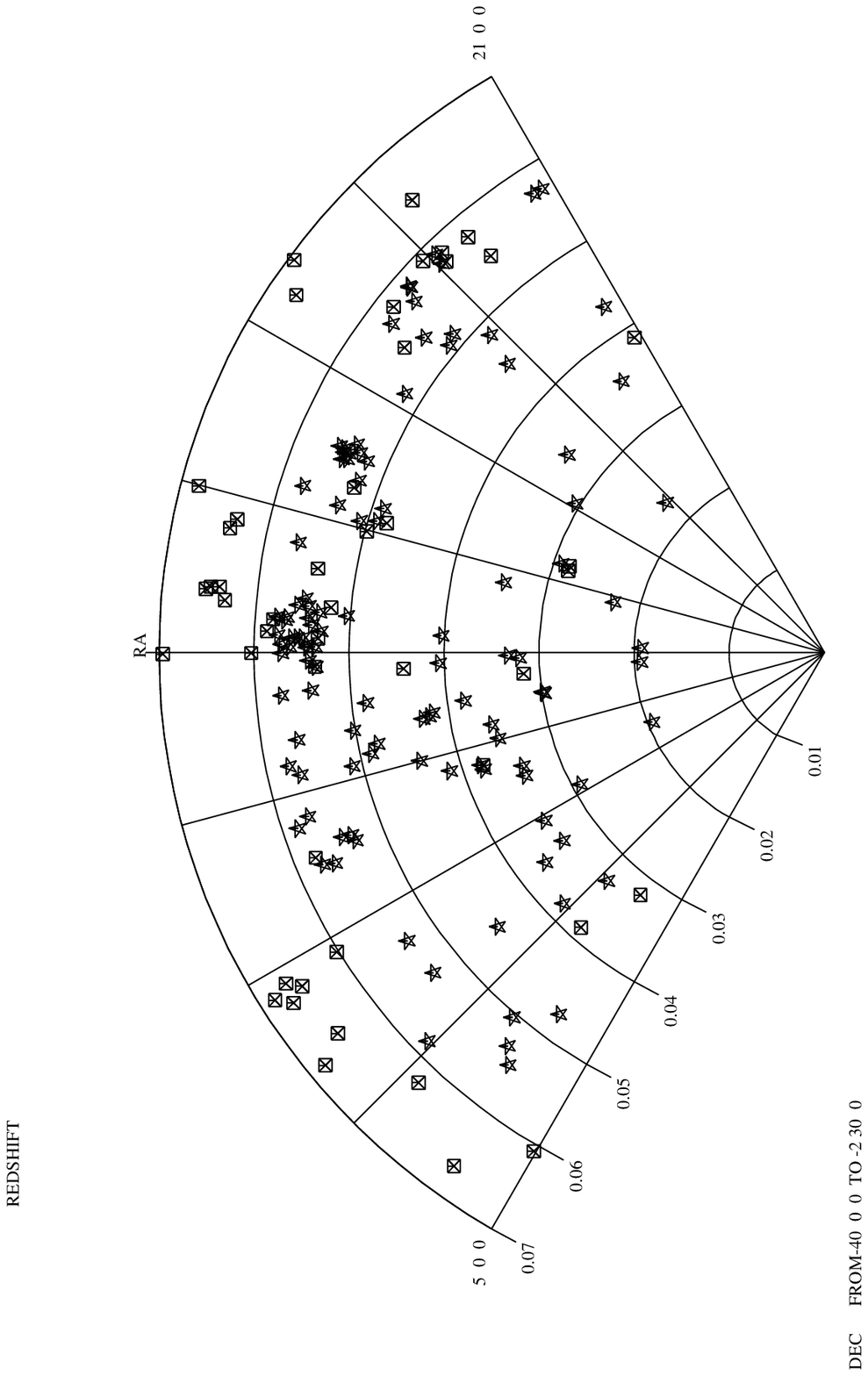] 
{ Spatial distribution of VLGs and ACO cluster. 
The cone diagram includes the $M_B \le -21$ galaxy subsample 
(stars), and the ACO clusters in the same $b_{II}$ and $\delta$ 
limits up to a redshift $z = 0.07$ (triangles).
}

\figcaption[]{ 
Four examples of distant VLGs with distorted morphology, faint 
satellite galaxies and/or companions or IG. 
Fields are about $11 \times 11$ arcminutes (corresponding 
to $\sim 2 \times 2$ \h~ at the typical distances of these galaxies).
}

\figcaption[]{ 
Mosaic showing the images of
the 13 nearby VLGs ($M \le -21$ and $V \le 10000$ km/s). 
Fields are $15 \times 15$ square arcminutes.
}

\figcaption[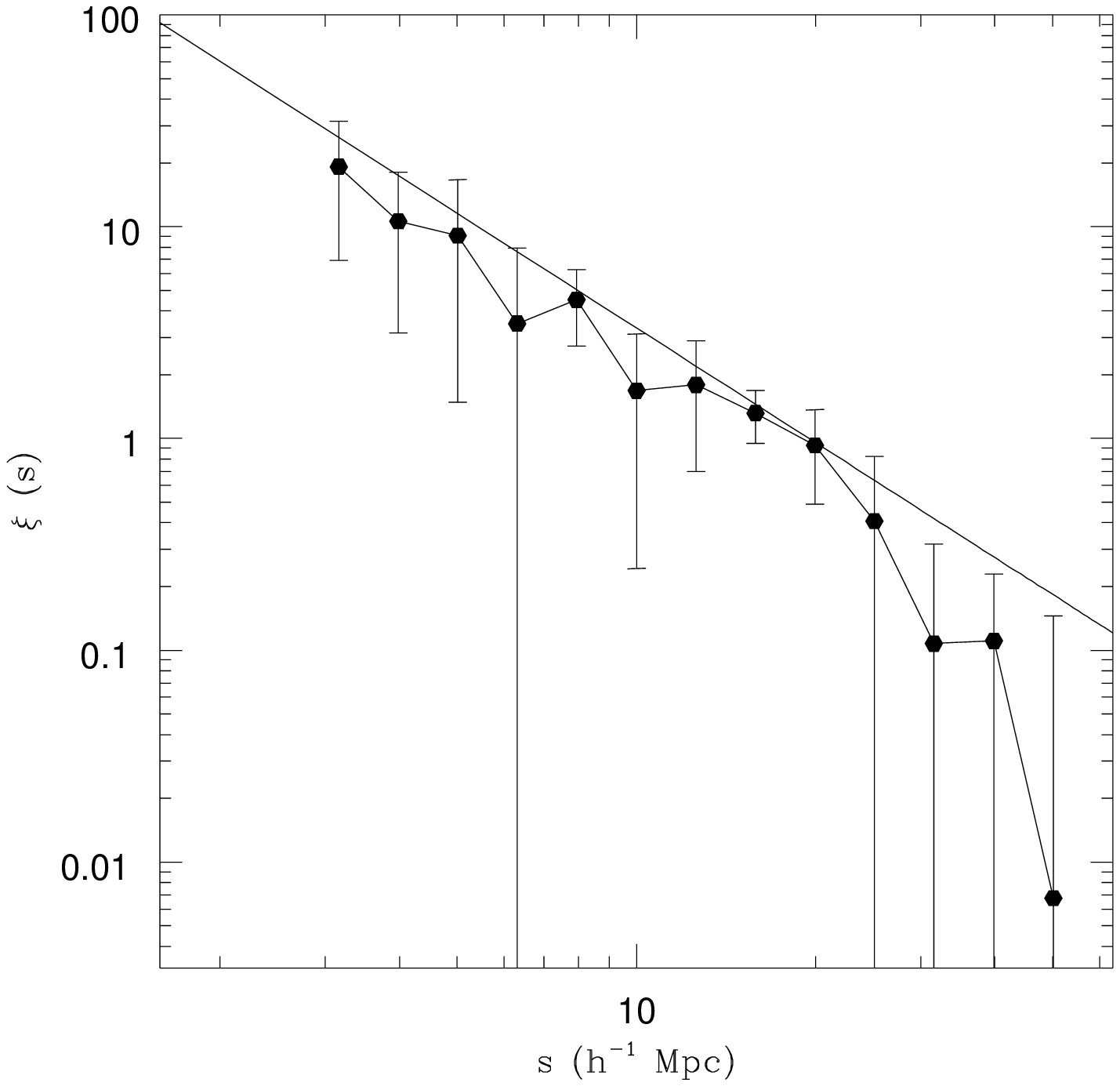]{ 
Comparison of the VLG and cluster autocorrelation functions.
Filled hexagons: volume--limited sample, $D_{lim}=168.5$ \h, for galaxies
with \MB $ \le -21$.
Solid straight line: fit for $R \ge 0$ ACO clusters
(Cappi \& Maurogordato 1992).
}

\figcaption[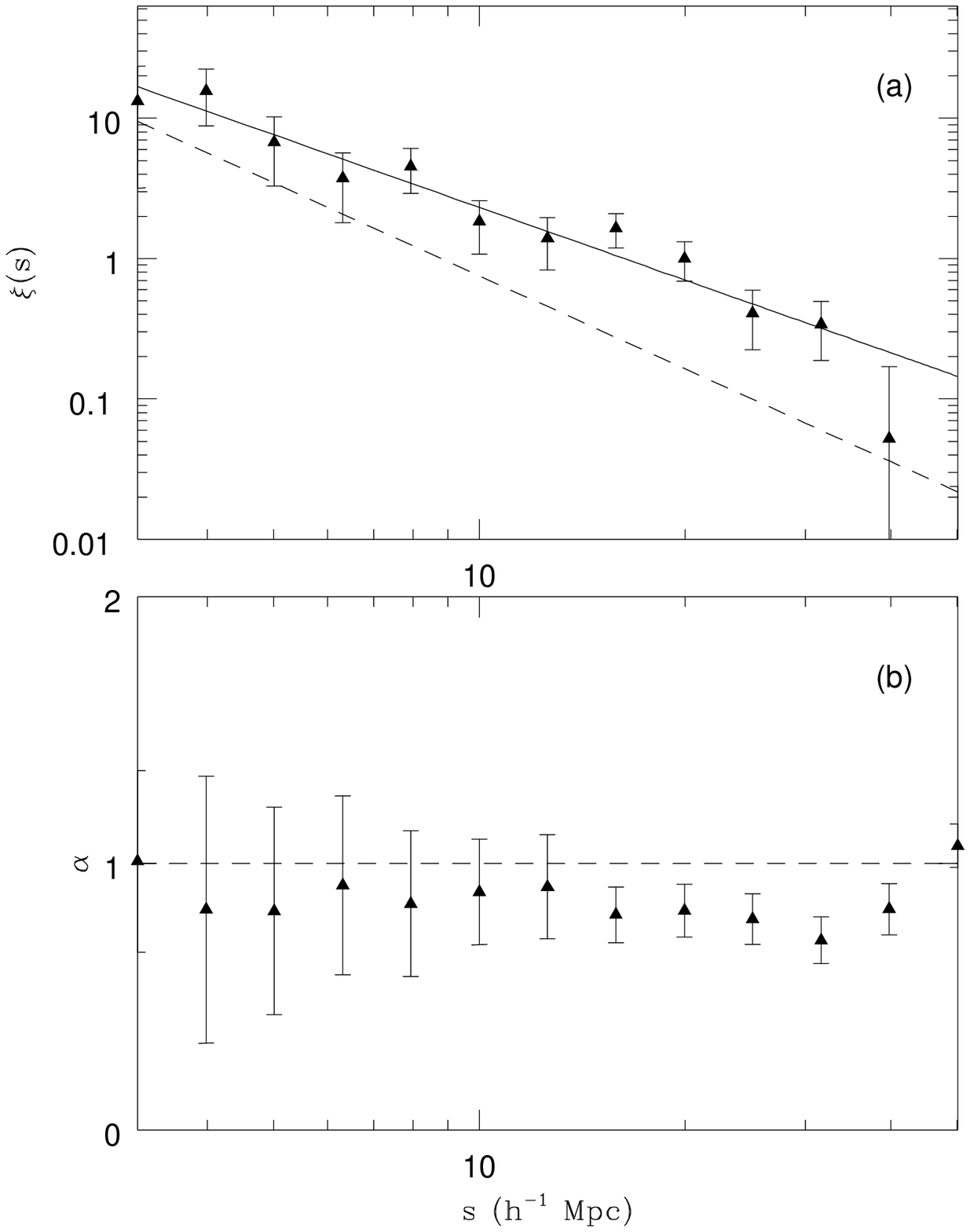]{ 
a) Galaxy--cluster cross--correlation function for galaxies 
brighter than \MB $ = -21$ (subsample limited at $D_{lim} = 168.5$ \h)
The solid line gives the best fit to the data. The dashed line is the
fit to the standard galaxy--cluster cross--correlation by Lilje \&
Efstathiou (1988).
b) The ratio $\alpha = J_3 (gc) / J_3 (cc)$ as a function of separation.
}

\figcaption[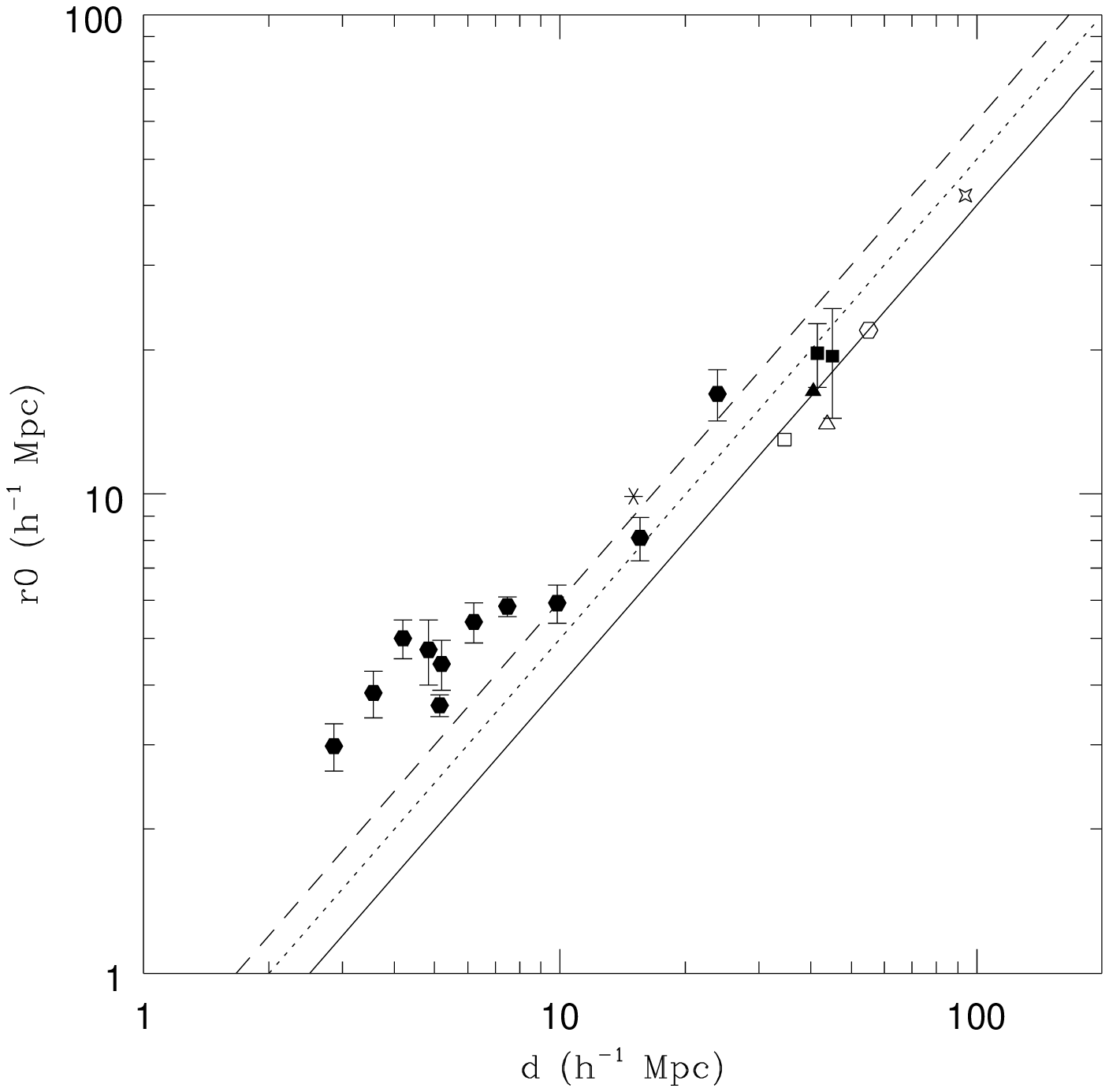]{ 
The correlation length
$r_0$ as a function of the mean interparticle distance $d_i = n^{-1/3}$.
Filled hexagons: galaxy subsamples of the SSRS2 from \MB $=-16$ to \MB $=-21$.
Asterisk: CfA2+SSRS2 groups from Girardi et al. (1998).
Filled triangle: radiogalaxies (from Bahcall \& Chokshi 1992).
Filled squares: 2 northern and southern ACO cluster subsamples with
$R \ge 0$ as defined in Cappi \& Maurogordato 1992.
Open hexagon: $R \ge 1$ Abell clusters.
Star: very rich Abell clusters ($R \ge 2$; see Bahcall \& West 1992).
Lines represent the relation $A_i \sim (\alpha d_i)^{1.8}$.
Solid line: $\alpha = 0.4$; dotted line: $\alpha = 0.5$;
dashed line: $\alpha = 0.6$.
}
\vfill
\newpage

%
%

\topmargin -2.0cm
\begin{figure*}[h]
\figurenum{1}
\caption[]{}
\label{fig:1}
\epsfysize=15.0cm \epsfbox{vlg.fig1.ps}
\end{figure*}

\topmargin -2.0cm
\begin{figure*}[h]
\figurenum{2}
\caption[]{}
\label{fig:2}
\epsfysize=15.0cm \epsfbox{vlg.fig2.ps}
\end{figure*}

\newpage

\topmargin -6.0cm
\begin{figure*}[h]
\figurenum{3}
\caption[]{}
\label{fig:3}
\epsfysize=25.0cm \epsfbox{vlg.fig3.ps}
\end{figure*}
\vfill

%
%
\topmargin -2.0cm
\begin{figure*}[h]
\figurenum{6}
\caption[]{}
\label{fig:test}
\epsfbox{vlg.fig6.ps}
\end{figure*}

\newpage

%
%
\topmargin -2.0cm
\begin{figure*}
\figurenum{7}
\caption[]{}
\label{fig:cross}
\epsfbox{vlg.fig7.ps}
\end{figure*}

\clearpage
%
%
\topmargin -2.0cm
\begin{figure*}
\figurenum{8}
\caption[]{}
\label{fig:universal}
\epsfbox{vlg.fig8.ps}
\end{figure*}

\end{document}


\begin{center}
 \begin{deluxetable}{rllrrllrlrl} \tablenum{1} 
 \tablecaption{\label{tab:1}}
 \tablewidth{0pt}
 \tablehead{
 \colhead{No.} & \colhead{Catalogue} &  \colhead{Other} & \colhead{RA} & \colhead{Dec} &  \colhead{$m_B$} &  \colhead{B-R} & \colhead{log$_{10} L_{60}$} &  \colhead{Type} &  \colhead{$V_\odot$} & \colhead{Notes} \nl & \colhead{Name} & \colhead{Name}  & & & & & & \colhead{(km/s)} & }
 \startdata

   1 & VLG 0006-3408 & ESO 349- G 032  &  00 06 02.2 & -34 08 11 & 13.19 &  1.51 &  9.62 & SAB(rs)b     &  6813 & ir         \nl
   2 & VLG 0010-2429 & ESO 472- G 020  &  00 10 30.7 & -24 29 34 & 14.00 &  2.24 & 10.00 & SA(r)c       & 10269 & ir         \nl
   3 & VLG 0012-2458 & ESO 473- G 003  &  00 12 33.8 & -24 58 11 & 15.13 &  0.73 & 10.27 & Sb           & 16867 & ir         \nl
   4 & VLG 0036-1541 &                 &  00 36 08.0 & -15 41 39 & 14.87 &   & 10.41 &  5           & 16352 & ir         \nl
   5 & VLG 0039-1708 &                 &  00 39 15.2 & -17 08 05 & 14.47 &   &    & SB(s)d       & 16239 &            \nl
   6 & VLG 0039-0934 &                 &  00 39 18.6 & -09 34 38 & 13.98 &   &    & cD/SB(r)     & 16536 & rs         \nl
   7 & VLG 0041-2522 & ESO 474- G 020  &  00 41 46.2 & -25 22 58 & 15.13 &  1.12 &    & Sb           & 15921 &            \nl
   8 & VLG 0042-2839 &                 &  00 42 22.9 & -28 39 47 & 14.98 &   & 10.32 & SA(r)a       & 15061 & ir         \nl
   9 & VLG 0044-2438 & ESO 474- G 026  &  00 44 39.8 & -24 38 37 & 15.19 &  0.97 & 10.52 & SA(s)a       & 16246 & ir         \nl
  10 & VLG 0044-0705 &                 &  00 44 51.3 & -07 05 03 & 15.21 &   &    & -5           & 16843 &            \nl
  11 & VLG 0044-1133 &                 &  00 44 53.5 & -11 33 43 & 15.00 &   &    &  6           & 17157 &            \nl
  12 & VLG 0046-2350 & ESO 474- G 031  &  00 46 15.5 & -23 50 03 & 15.36 &  1.60 &    & S0           & 16987 &            \nl
  13 & VLG 0047-1457 &                 &  00 47 30.9 & -14 57 16 & 15.12 &   &    & -5           & 16086 &            \nl
  14 & VLG 0049-2735 &                 &  00 49 48.9 & -27 35 59 & 14.50 &   &    & Sbc          & 12045 &            \nl
  15 & VLG 0050-2314 & ESO 474- G 036  &  00 50 18.7 & -23 14 47 & 15.20 &  1.44 &    & S0:          & 15846 &            \nl
  16 & VLG 0052-3535 & ESO 351- G 021  &  00 52 37.1 & -35 35 33 & 15.21 &  1.73 &    & SA0-         & 17254 &            \nl
  17 & VLG 0053-1015 &                 &  00 53 20.9 & -10 15 21 & 15.14 &   &    & -2           & 16352 &            \nl
  18 & VLG 0053-3740 & ESO 295-IG 022  &  00 53 24.2 & -37 40 43 & 15.23 &   &    & E            & 16657 &            \nl
  19 & VLG 0053-1337 &                 &  00 53 34.3 & -13 37 43 & 15.18 &   &    &  5           & 16710 &            \nl
  20 & VLG 0054-1011 &                 &  00 54 11.6 & -10 11 05 & 12.50 &   &  9.94 & SAB(r)c      &  5666 &            \nl
  21 & VLG 0056-1644 &                 &  00 56 22.4 & -16 44 19 & 14.70 &   &    & -2           & 16226 &            \nl
  22 & VLG 0056-2834 &                 &  00 56 27.6 & -28 34 23 & 15.31 &   &    & S0           & 17079 &            \nl
  23 & VLG 0057-1415 &                 &  00 57 29.5 & -14 15 50 & 15.20 &   &    & -2           & 15972 &            \nl
  24 & VLG 0057-2145 & ESO 541- G 007  &  00 57 34.9 & -21 45 28 & 15.09 &  1.59 &    & E-S0         & 16805 &            \nl
  25 & VLG 0057-1534 &                 &  00 57 46.8 & -15 34 06 & 15.28 &   &    & -2           & 16235 & rs         \nl
  26 & VLG 0100-2209 & ESO 541- G 013  &  01 00 15.1 & -22 09 02 & 14.84 &  1.60 &    & cD;E+3pec    & 17073 & rs         \nl
  27 & VLG 0101-3523 & ESO 352- G 001  &  01 01 57.8 & -35 23 21 & 14.04 &  1.08 & 10.36 & SB(r)bc pec  &  9991 & ir         \nl
  28 & VLG 0103-1916 &                 &  01 03 37.1 & -19 16 19 & 15.19 &   &    & -2           & 16194 &            \nl
  29 & VLG 0103-3026 & ESO 412- G 00   &  01 03 49.3 & -30 26 43 & 13.72 &  1.10 & 10.44 & SB(r)c       &  9620 & ir         \nl
  30 & VLG 0106-1606 &                 &  01 06 19.4 & -16 06 38 & 14.46 &   &  9.74 & SA(rs)a      & 12073 & ir         \nl
  31 & VLG 0106-1540 &                 &  01 06 22.6 & -15 40 24 & 14.36 &   &    & D            & 15967 &            \nl
  32 & VLG 0111-3200 & ESO 412- G 018  &  01 11 26.4 & -32 00 43 & 12.59 &  1.57 &    & SAB(rs)0-    &  5803 &            \nl
  33 & VLG 0116-2547 &                 &  01 16 49.8 & -25 47 35 & 14.62 &   & 10.75 & SA(s)a       & 16188 & ir         \nl
  34 & VLG 0117-0530 &                 &  01 17 58.9 & -05 30 59 & 14.96 &   &    &  5           & 17040 &            \nl
  35 & VLG 0125-0456 &                 &  01 25 01.7 & -04 56 31 & 14.96 &   &    & S0           & 14407 &            \nl
  36 & VLG 0130-1504 &                 &  01 30 24.8 & -15 04 16 & 14.33 &   & 10.05 &  5           & 11387 & ir         \nl
  37 & VLG 0131-3644 & ESO 353- G 015  &  01 31 43.5 & -36 44 57 & 13.16 &  1.57 & 10.25 & SA0+pec sp   &  8985 &            \nl
  38 & VLG 0132-3644 & ESO 353- G 023  &  01 32 52.4 & -36 44 44 & 13.88 &  1.64 &    & E+:          &  8955 &            \nl
  39 & VLG 0134-0927 &                 &  01 34 46.2 & -09 27 08 & 14.33 &   &    & SA0-         & 12394 & rs         \nl
  40 & VLG 0136-2849 &                 &  01 36 41.1 & -28 49 33 & 14.64 &   &    & S            & 12628 &            \nl
  41 & VLG 0137-0929 &                 &  01 37 27.0 & -09 29 43 & 14.50 &   &    & SB(r)0?      & 12741 &            \nl
  42 & VLG 0137-1506 &                 &  01 37 33.4 & -15 06 56 & 14.93 &   & 10.36 &  5           & 14973 & ir         \nl
  43 & VLG 0137-2812 & ESO 413- G 015  &  01 37 38.2 & -28 12 34 & 15.28 &  0.99 &    & Sb           & 16825 &            \nl
  44 & VLG 0146-1238 &                 &  01 46 05.0 & -12 38 09 & 14.92 &   & 10.27 &  5           & 14301 & ir         \nl
  45 & VLG 0147-2907 &                 &  01 47 57.4 & -29 07 10 & 14.95 &   & 10.01 & S            & 17257 & ir         \nl
  46 & VLG 0148-0418 &                 &  01 48 47.5 & -04 18 06 & 14.23 &   & 10.33 &  5           & 10607 &            \nl
  47 & VLG 0150-1901 & ESO 543- G 022  &  01 50 16.8 & -19 01 39 & 14.92 &   &    & S?           & 14580 &            \nl
  48 & VLG 0152-2823 & ESO 414-IG 009  &  01 52 32.0 & -28 23 38 & 15.17 &  1.12 & 10.37 &  4           & 16945 & ir         \nl
  49 & VLG 0154-2017 & ESO 543- G 026  &  01 54 10.8 & -20 17 33 & 15.08 &   &  9.84 & Sb:          & 15252 & ir         \nl
  50 & VLG 0158-3327 & ESO 354- G 029  &  01 58 58.6 & -33 27 02 & 14.09 &  1.20 &  9.83 & Sc           & 10644 & ir         \nl
  51 & VLG 0159-2200 & ESO 544- G 007  &  01 59 56.6 & -22 00 12 & 14.47 &  1.45 &  9.89 & S?           & 13161 & ir         \nl
  52 & VLG 0210-2242 & ESO 478- G 013  &  02 10 11.5 & -22 42 20 & 14.13 &  0.68 & 10.44 & SB(rs)c?     & 12356 & ir         \nl
  53 & VLG 0210-2641 &                 &  02 10 33.3 & -26 41 35 & 15.16 &   &    & S0           & 17067 &            \nl
  54 & VLG 0212-2504 & ESO 478- G 015  &  02 12 32.5 & -25 04 59 & 14.41 &  1.17 &    & R('2SB(s)b   & 11394 &            \nl
  55 & VLG 0213-2416 &                 &  02 13 57.5 & -24 16 37 & 15.37 &   & 10.05 & S            & 17483 & ir         \nl
  56 & VLG 0215-2336 & ESO 478- G 022  &  02 15 36.8 & -23 36 53 & 14.38 &  1.13 & 10.19 & Sa?          & 11332 & ir         \nl
  57 & VLG 0222-1922 & ESO 545- G 013  &  02 22 20.5 & -19 22 02 & 13.59 &  1.15 & 10.33 & SAB(s)c Sy1. & 10132 & ir,an      \nl
  58 & VLG 0223-2411 & ESO 479- G 003  &  02 23 44.6 & -24 11 07 & 15.24 &  1.80 &    & S0           & 16266 &            \nl
  59 & VLG 0224-2338 &                 &  02 24 19.8 & -23 38 28 & 14.76 &   &    & S0           & 15982 &            \nl
  60 & VLG 0227-1715 &                 &  02 27 17.4 & -17 15 57 & 15.09 &   &    &  3           & 15872 & rs         \nl
  61 & VLG 0228-0309 &                 &  02 28 11.1 & -03 09 38 & 12.73 &   & 10.39 & SB(rs)c      &  5738 &            \nl
  62 & VLG 0229-2006 & ESO 545- G 030  &  02 29 27.8 & -20 06 18 & 13.82 &  1.61 &    & SA0 pec      & 10165 &            \nl
  63 & VLG 0231-2115 & ESO 545- G 036  &  02 31 38.0 & -21 15 43 & 15.36 &  1.55 &    & R'?SA(s)a    & 17137 &            \nl
  64 & VLG 0233-2544 & ESO 479- G 015  &  02 33 07.8 & -25 44 43 & 15.10 &  1.34 &    & SB(rs)bc     & 16799 &            \nl
  65 & VLG 0253-1424 &                 &  02 53 22.1 & -14 24 34 & 13.76 &   &    & SA0-:        &  8703 & rs         \nl
  66 & VLG 0304-0955 &                 &  03 04 03.1 & -09 55 24 & 14.20 &   &    & SA0- pec:    & 10239 &            \nl
  67 & VLG 0318-3842 & ESO 301-IG 006  &  03 18 39.0 & -38 42 52 & 15.08 &  1.40 &    &  1           & 16068 &            \nl
  68 & VLG 0322-0621 &                 &  03 22 43.5 & -06 21 22 & 13.97 &   & 10.57 & S Sy1        & 10107 & ir,an      \nl
  69 & VLG 0327-2856 & ESO 418- G 007  &  03 27 52.0 & -28 56 38 & 13.95 &  1.49 &  9.86 & SB(s)b pec   & 11052 & ir         \nl
  70 & VLG 0337-3341 & ESO 358-IG 047  &  03 37 14.3 & -33 41 32 & 14.84 &  0.56 & 10.61 & S pec        & 16008 & ir         \nl
  71 & VLG 0340-3811 & ESO 301-IG 025  &  03 40 09.3 & -38 11 28 & 14.76 &  0.91 & 10.45 & SB(s)I-II    & 13542 & ir         \nl
  72 & VLG 0355-1921 & ESO 549- G 042  &  03 55 24.5 & -19 21 34 & 14.36 &  1.64 &    & SB(r)b       & 11449 &            \nl
  73 & VLG 0357-3855 & ESO 302- G 025  &  03 57 39.8 & -38 55 53 & 14.95 &  1.47 &    & S...         & 17618 &            \nl
  74 & VLG 0405-0857 &                 &  04 05 43.3 & -08 57 53 & 13.95 &   & 10.60 & 21           &  9928 & ir         \nl
  75 & VLG 0417-3917 & ESO 303-IG 011  &  04 17 55.0 & -39 17 30 & 14.93 &   &  9.96 & Irr          & 15255 & ir         \nl
  76 & VLG 0424-3440 & ESO 360- G 009  &  04 24 46.2 & -34 40 34 & 15.18 &  2.00 &    & SA(rs)0-?    & 16031 &            \nl
  77 & VLG 0430-3030 & ESO 421- G 009  &  04 30 19.9 & -30 30 15 & 14.98 &  1.40 & 10.56 & S(r)a Lin    & 16481 & ir,an      \nl
  78 & VLG 0435-3944 & ESO 304- G 006  &  04 35 07.5 & -39 44 45 & 14.71 &  0.91 &    & SAB(rs)b     & 14283 &            \nl
  79 & VLG 2105-3846 & ESO 342- G 007  &  21 05 26.4 & -38 46 06 & 15.30 &  0.41 &    & Sa-b         & 17162 &            \nl
  80 & VLG 2109-3919 & ESO 342- G 020  &  21 09 25.3 & -39 19 56 & 15.26 &  1.34 &    & Sc           & 17159 &            \nl
  81 & VLG 2109-3804 & ESO 342- G 022  &  21 09 47.4 & -38 04 33 & 14.50 &  0.92 & 10.48 & Sb           & 12944 & ir         \nl
  82 & VLG 2126-2124 & ESO 599- G 016  &  21 26 47.6 & -21 24 54 & 14.04 &  2.15 &    & SA0-:        & 10608 & rs         \nl
  83 & VLG 2157-1926 &                 &  21 57 19.8 & -19 26 41 & 15.48 &   &    & E            & 17473 &            \nl
  84 & VLG 2158-1507 &                 &  21 58 02.2 & -15 07 27 & 14.92 &   &    &  5           & 17096 &            \nl
  85 & VLG 2205-1919 & ESO 601- G 021  &  22 05 37.5 & -19 19 06 & 14.67 &  1.43 &  9.87 & SB(r)bc      & 14448 & ir         \nl
  86 & VLG 2206-2803 & ESO 467- G 012  &  22 06 17.2 & -28 03 20 & 13.05 &  1.08 & 10.11 & SB(s)bc p:Sy &  6832 & ir,an      \nl
  87 & VLG 2210-1653 &                 &  22 10 22.9 & -16 53 21 & 14.72 &   & 10.11 &  3           & 13403 & ir         \nl
  88 & VLG 2214-3711 &                 &  22 14 10.2 & -37 11 52 & 15.08 &   &    &  5           & 17471 &            \nl
  89 & VLG 2214-0455 &                 &  22 14 20.5 & -04 55 08 & 15.42 &   & 10.44 & -5           & 17259 & ir         \nl
  90 & VLG 2217-2033 &                 &  22 17 18.9 & -20 33 59 & 14.79 &   & 10.38 &  5           & 15338 & ir         \nl
  91 & VLG 2217-3256 & ESO 405- G 009  &  22 17 33.4 & -32 56 17 & 15.10 &  1.08 &    & Sa           & 16992 &            \nl
  92 & VLG 2222-1831 &                 &  22 22 37.1 & -18 31 24 & 15.05 &   & 10.17 &  5           & 15188 & ir         \nl
  93 & VLG 2227-3046 & ESO 468- G 003  &  22 27 01.4 & -30 46 51 & 14.73 &  0.97 &    & Sb           & 16021 &            \nl
  94 & VLG 2229-2539 & ESO 533- G 044  &  22 29 22.3 & -25 39 20 & 13.92 &  1.47 &    & SAB(rs)p     & 10157 &            \nl
  95 & VLG 2231-3939 &                 &  22 31 00.2 & -39 39 22 & 14.98 &   &  9.98 & S            & 17150 & ir         \nl
  96 & VLG 2253-3449 & ESO 406- G 022  &  22 53 05.7 & -34 49 19 & 14.85 &  0.46 &    & N            & 15519 &            \nl
  97 & VLG 2256-2547 & ESO 535- G 001  &  22 56 19.0 & -25 47 47 & 13.84 &   &  9.85 & (R:)SA(rs)a  &  9091 & ir         \nl
  98 & VLG 2323-3809 &                 &  23 23 52.6 & -38 09 29 & 14.92 &   &    & S            & 16135 &            \nl
  99 & VLG 2326-3811 &                 &  23 26 51.3 & -38 11 18 & 15.06 &   &    & S            & 15933 &            \nl
 100 & VLG 2327-3924 & ESO 347- G 019  &  23 27 34.6 & -39 24 13 & 15.21 &  1.43 &    & SB(r)ab      & 16611 &            \nl
 101 & VLG 2327-3933 &                 &  23 27 48.9 & -39 33 26 & 14.79 &   &    & E            & 16451 & rs         \nl
 102 & VLG 2329-3933 &                 &  23 29 07.9 & -39 33 39 & 15.05 &   &    & -2           & 15644 &            \nl
 103 & VLG 2329-3419 & ESO 408- G 008  &  23 29 11.2 & -34 19 54 & 15.06 &  1.77 &    & SAB(rs)      & 16234 &            \nl
 104 & VLG 2329-3949 &                 &  23 29 22.9 & -39 49 52 & 15.13 &   &    & S            & 16413 &            \nl
 105 & VLG 2331-3837 &                 &  23 31 46.2 & -38 37 15 & 15.16 &   &    & (R')SB(r)b   & 16255 &            \nl
 106 & VLG 2332-3823 & ESO 347- G 024  &  23 32 32.3 & -38 23 54 & 15.08 &  1.43 &    & S0-a         & 16387 &            \nl
 107 & VLG 2338-2935 &                 &  23 38 52.8 & -29 35 53 & 15.05 &   &    & E            & 15608 & rs         \nl
 108 & VLG 2345-2825 & ESO 471- G 019  &  23 45 08.3 & -28 25 03 & 13.78? &  1.43 &    & cD;E+4 pec   &  8695 &            \nl
 109 & VLG 2347-2917 &                 &  23 47 38.1 & -29 17 13 & 14.94 &   &    & E            & 14580 &            \nl
 110 & VLG 2348-3443 & ESO 349- G 001  &  23 48 54.5 & -34 43 50 & 14.52 &  1.81 &    & E+1          & 17239 &            \nl
 111 & VLG 2352-3452 & ESO 349- G 005  &  23 52 22.6 & -34 52 47 & 14.76 &  1.19 &    & Sc           & 16002 &            \nl
 112 & VLG 2354-3502 & ESO 349- G 010  &  23 54 25.9 & -35 02 16 & 14.29 &  1.70 &    & cD;E+4;Lin   & 14696 & rs,an      \nl
 113 & VLG 2356-2104 & ESO 538- G 014  &  23 56 34.3 & -21 04 35 & 15.02 &  1.14 &    & Sa:          & 15140 &            \nl

 \enddata
 \end{deluxetable}
 \end{center}


\begin{center}
 \begin{deluxetable}{rlllrr} \tablenum{2} 
 \tablecaption{\label{tab:2}}
 \tablewidth{0pt}
 \tablehead{
 \colhead{No}. & \colhead{Catalogue} & \colhead{System type} &  
  \colhead{System name} & \colhead{$\sigma$} &  \colhead{$N_g$} \nl
 & \colhead{name} & & & \colhead{(km/s)} & } 
 \startdata

   1 &  VLG 0006-3408 & group candidate       & SSRS2\tablenotemark{a}      & &  \nl
   4 &  VLG 0036-1541 & triplet; interacting  & IC 0038                     & &  \nl
   6 &  VLG 0039-0934 & rich cluster; cD      & A0085                       & &  \nl
   7 &  VLG 0041-2522 & binary                & ESO 474- G 020              & &  \nl
  14 &  VLG 0049-2735 & triplet               & AM 0049-273                 & &  \nl
  17 &  VLG 0053-1015 & rich cluster, binary  & A0117                       & &  \nl
  18 &  VLG 0053-3740 & binary                & S0102                       & &  \nl
  20 &  VLG 0054-1011 & group candidate       & SSRS2\tablenotemark{a}                   &  907 &    9 \nl
  21 &  VLG 0056-1644 & binary                & NGC0333A/B                  & &  \nl
  22 &  VLG 0056-2834 & group, binary         & SCG55                       & &  \nl
  23 &  VLG 0057-1415 & rich cluster          & A0126                       & &  \nl
  26 &  VLG 0100-2209 & rich cluster; cD pec  & A0133                       & &  \nl
  30 &  VLG 0106-1606 & group                 & SSRS2\tablenotemark{b}                & &  \nl
  31 &  VLG 0106-1540 & rich cluster          & A0151                       & &  \nl
  32 &  VLG 0111-3200 & poor cluster          & S0141                       &  292 &   17 \nl
  35 &  VLG 0125-0456 & compact group         & HCG12                       & &  \nl
  36 &  VLG 0130-1504 & binary; interacting    & IC0141                      & &  \nl
  37 &  VLG 0131-3644 & group                 & SSRS2\tablenotemark{b}                   &   68 &    5 \nl
  38 &  VLG 0132-3644 & group                 & SSRS2\tablenotemark{b}                   &  150 &    7 \nl
  46 &  VLG 0148-0418 & binary; interacting    & NGC0702                     & &  \nl
  47 &  VLG 0150-1901 & triplet               & ESO 543- G 022              & &  \nl
  48 &  VLG 0152-2823 & binary                & ESO 414-IG 009              & &  \nl
  49 &  VLG 0154-2017 & group                 & SCG14                       & &  \nl
  51 &  VLG 0159-2200 & group                 & VV583                       & &  \nl
  52 &  VLG 0210-2242 & group                 & SSRS2\tablenotemark{b}                   & &  \nl
  54 &  VLG 0212-2504 & rich cluster, triplet & A2995                       & &  \nl
  56 &  VLG 0215-2336 & group                 & SSRS2\tablenotemark{b}                   &  877 &    9 \nl
  57 &  VLG 0222-1922 & group Seyfert         & SSRS2\tablenotemark{b}                   &  146 &    6 \nl
  59 &  VLG 0224-2338 & poor cluster; brightest member & S0263               & &  \nl
  60 &  VLG 0227-1715 & poor cluster, binary  & S0268                       & &  \nl
  62 &  VLG 0229-2006 & group                 & SSRS2\tablenotemark{b}                   & &  \nl
  65 &  VLG 0253-1424 & group candidate; extended X source & SSRS2\tablenotemark{a} & &  \nl
  67 &  VLG 0318-3842 & binary                & ESO 301-IG 006              & &  \nl
  68 &  VLG 0322-0621 & group                 & SSRS2\tablenotemark{b}                   & &  \nl
  69 &  VLG 0327-2856 & binary; interacting    & ESO 418- G 007              & &  \nl
  70 &  VLG 0337-3341 & binary; interacting    & ESO 358-IG 047              & &  \nl
  71 &  VLG 0340-3811 & group; interacting     & AM 0340-381                 & &  \nl
  74 &  VLG 0405-0857 & binary                & IRAS 04057-0857             & &  \nl
  75 &  VLG 0417-3917 & group; interacting     & ESO 303-IG 011              & &  \nl
  80 &  VLG 2109-3919 & binary                & ESO 342- G 020              & &  \nl
  86 &  VLG 2206-2803 & compact group; Seyfert & HCG91 (VV700)              &  180 &   11 \nl
  88 &  VLG 2214-3711 & group                 & AM 2214-371                 & &  \nl
  91 &  VLG 2217-3256 & triplet               & AM 2217-325                 & &  \nl
  94 &  VLG 2229-2539 & poor cluster          & EDCC155                     &   57 &    3 \nl
  95 &  VLG 2231-3939 & poor cluster          & S1036                       & &  \nl
  97 &  VLG 2256-2547 & group candidate       & SSRS2\tablenotemark{a}                   & &  \nl
 100 &  VLG 2327-3924 & rich cluster          & A4008                       &  430 &    4 \nl
 101 &  VLG 2327-3933 & rich cluster; dumbbell system & A4008               & &  \nl
 102 &  VLG 2329-3933 & rich cluster          & A4008                       & &  \nl
 103 &  VLG 2329-3419 & poor cluster          & EDCC312                     & &  \nl
 104 &  VLG 2329-3949 & rich cluster          & A4008                       & &  \nl
 107 &  VLG 2338-2935 & poor cluster          & EDCC332                     & &  \nl
 108 &  VLG 2345-2825 & rich cluster; cD pec  & A4038                       &  307 &   16 \nl
 112 &  VLG 2354-3502 & rich cluster; cD , Liner & A4059                    & &  \nl

 \enddata
\tablenotetext{a}{SSRS2 group candidate found in this work.}
\tablenotetext{b}{SSRS2 group listed in the Girardi et al. catalog (1998).}
 \end{deluxetable}
 \end{center}